\documentclass[twocolumn,prl,showpacs,floatfix,superscriptaddress]{revtex4}
\usepackage{graphicx}
\usepackage{amsmath}

\usepackage{bm}
\begin{document}

\title{Equivalence of operator-splitting schemes for the
integration of the Langevin equation}

\author {H. K. Lee}
\affiliation{School of Physics, Korea Institute for Advanced Study, Seoul
130-722, Korea} \author{C. Kwon} \affiliation{Department of Physics, Myongji
University, Yongin, Gyeonggi-Do 449-728, Korea} \author{Hyunggyu Park}
\affiliation{School of Physics, Korea Institute for Advanced Study, Seoul
130-722, Korea}
\date{\today}

\begin{abstract}
We investigate the equivalence of different operator-splitting
schemes for the integration of the Langevin equation. We consider a
specific problem, so called the directed percolation process, which
can be extended to a wider class of problems. We first give a
compact mathematical description of the operator-splitting method
and introduce two typical splitting schemes that will be useful in
numerical studies. We show that the two schemes are essentially
equivalent through the map that turns out to be an automorphism. An
associated equivalent class of operator-splitting integrations is
also defined by generalizing the specified equivalence.
\end{abstract}
 \maketitle

Several kinds of models for lattice-based dynamic processes have been studied
for decades to understand the characteristics of non-equilibrium systems,
especially, focused on the critical phenomena. To mention a few, directed
percolation (DP)~\cite{dp-geometric,Obukov,Jans,Grass,CS,DW-stochasticCA},
contact process~\cite{ContactProcess,GrassTorre},
and catalytic reactions~\cite{ZGB-CR} are such
examples. 
In these studies, it was found that, in spite of diversity in microscopic
details, various models exhibit critical phenomena that are essentially
identical to those for the absorbing phase transition in the DP model and the
DP universality class was inferred for
them~\cite{HinrichsenAP2000,OdorRMP2004}.

In order to explain the DP universality class from the unified point of view,
so called the DP Langevin equation is proposed~\cite{Jans,Grass} as
\begin{equation}
\partial_t \rho = a\rho - b\rho^2
+ D \nabla^2 \rho + \sigma \sqrt{\rho}~\xi, \label{dple}
\end{equation}
where $\rho=\rho(\mathbf{r},t)$ is a non-negative field variable for
concentration and $\xi$ is a white noise with zero mean satisfying $\langle
\xi(\mathbf{r},t) \xi(\mathbf{r}',t') \rangle =
2\delta(\mathbf{r}-\mathbf{r}')\delta(t-t')$. The coefficient $a$ is the tuning
parameter for the phase transition, and $b,~D,~\sigma$ are positive constants.

The field theoretical approach to Eq.~(\ref{dple}), named as the {\it Reggeon
field theory}~\cite{Reggeon}, unveils the critical behavior of the DP
universality class for spatial dimensions comparable to or higher than the
critical one $d_c=4$. The mean field theory applies in high dimensions $(d>d_c
= 4)$ and the perturbative results are well established near and lower than
$d_c$ through the standard $\epsilon=d_c-d$ expansion. In lower dimensions, the
series expansion methods for lattice-based DP models provide the most precise
estimates for the critical scaling exponents~\cite{numericsLD}, which have been
also confirmed by extensive numerical simulations~\cite{HinrichsenAP2000}.

Attention has also been paid to the direct numerical integration of
Eq.~(\ref{dple}), especially, in the quantitative study of the
lower-dimensional cases~\cite{numericsonDPeq}. This seems to be a simple
numerical integration of the partial differential equation at a glance.
However, as long as the absorbing transition is concerned, one has to meet an
annoying block which is by no means easily tractable. The conventional Euler
integration technique using a discrete time interval $\Delta t$ may result in a
negative value of $\rho$ due to the uncontrolled random noise term, and then
any further sensible integration of the equation is impossible. In particular,
this nuisance may appear very easily for small $\rho$ where the noise term
$(\sim\sqrt{\rho})$ is dominant over the deterministic term $(\sim\rho)$. As
the absorbing critical behavior occurs in the $\rho\rightarrow 0$ limit, the
proper treatment to guarantee a small but positive value of $\rho$ is not only
of a technical interest, but also a critical issue in the numerical study of
the absorbing phase transition. Some numerical schemes have been put forward to
overcome such numerical fragility~\cite{numericsonDPeq}, but without much
success.

Recently, Dornic et al.~\cite{DornicPRL05} utilized the operator-splitting
method for integrating the Langevin equations (similar idea earlier in
Ref.~\cite{PechenikPRE99}) describing various kinds of absorbing critical
phenomena~\cite{MunozSeries}. In this method, the evolution in time is divided
into two parts, each of which is exactly solvable. The successive integration
of the two parts during $\Delta t$ is regarded as one-step time evolution of
Eq.~(\ref{dple}), getting exact as $\Delta t \to 0$. This method has a couple
of outstanding advantages over the preceding ones~\cite{numericsonDPeq}. As the
noise term is treated exactly, the non-negativity of $\rho$ is always
guaranteed. Therefore one can use a relatively large value for $\Delta t$ to
stay in the critical region where $\rho$ is small, which can also save the
computing time considerably.
An astonishing observation is that the critical behavior seems to
be fairly insensitive to the magnitude of $\Delta t$.  In fact,
even if $\Delta t =0.25$ in their paper, the critical point is
shifted only by $1\%$ from the extrapolated value in the limit of
$\Delta t \rightarrow 0$. They have reported many successful
applications of this method in various types of absorbing critical
phenomena.

It is a tough task to rigorously explain why the operator-splitting method with
a relatively large time interval yields a reliable result for the critical
behavior of a certain problem and find what is the criterion for this approach
to be valid. Our preliminary work based on the perturbative expansion in
$\Delta t$ reveals that the operator-splitting scheme renormalizes the given
coefficients ($a,\ b,\ D,\ \sigma$), generates higher order terms, and new
non-Gaussian noises. Furthermore, we find that it may be possible for the
renormalized and newly generated coefficients to change their signs, which
could produce a diverging solution, a first-order phase transition, or a
higher-order multicritical point. Unfortunately the coefficients are expressed
in terms of alternating infinite series, from which it seems hardly probable to
derive the validity criteria of $\Delta t$ for maintaining the DP critical
behavior.

In this paper, we present a compact mathematical description of
the operator-splitting method. We analytically show that some of
seemingly different splitting schemes are mathematically
equivalent in the sense that there is an exact transformation map
relating different splitting schemes. We hope that our result may
elucidate the structure of the $\Delta t$-dependent terms and
eventually help us understand the characteristics of the upper
bound for $\Delta t$ in general operator-splitting schemes.

First, we summarize the operator-splitting scheme specified in
Ref.~\cite{DornicPRL05} and then provide a compact mathematical
description for it. We next consider another scheme with a different
choice in splitting the dynamic process. Using the
Baker-Campbell-Hausdorff (BCH) formula,
we prove the equivalence of two
different schemes with the analytic expression of the parameter
transformation function.

In the numerical study, the embedding space is replaced with a mesh-like
lattice with lattice constant $\Delta x$. Then, at site $i$, Eq.~(\ref{dple})
becomes an ordinary differential equation for $\rho_i (t)$
\begin{equation}
\dot\rho_i = \tilde{a} \rho_i - b \rho_i^2 + {D\over (\Delta x)^2}
\sum_j\epsilon_{ij}\rho_j + \sigma \sqrt{\rho_i}~ \xi_i, \label{sode}
\end{equation}
where $\dot\rho_i$ is the time derivative of $\rho_i$, $\tilde{a}\equiv
a-2dD/(\Delta x)^2$, $\epsilon_{ij}=1$ for nearest neighbor sites $i$ and $j$
or $\epsilon_{ij}=0$ otherwise, and $d$ is the spatial dimension.
Eq.~(\ref{sode}) is a set of coupled equations where the dynamics at site $i$
is influenced by field variables $\rho_j$ at nearest neighbors. As an
additional approximation, we assume that $\rho_j$ is piecewise constant with an
initial value $\rho_j(t)$ during the integration from $t$ to $t+\Delta t$.
During this interval, Eq.~(\ref{sode}) then becomes decoupled with an effective
field $c_i\equiv D/(\Delta x)^2\sum_j\epsilon_{ij}\rho_j(t)$, which leads to
\begin{equation}
\dot\rho = c+\tilde{a} \rho - b \rho^2 + \sigma \sqrt{\rho}~ \xi,
\label{sode-1}
\end{equation}
where the subscript $i$ is dropped for simplicity.

The main idea of the operator-splitting method is to separate the right hand
side of Eq.~(\ref{sode-1}) into two parts, each of which can be treated
exactly. In particular, it is important to have or find the exact solution of
the Fokker-Planck (FP) equation associated with the part including the noise
term, guaranteeing the non-negativity of $\rho$. Dornic et
al.~\cite{DornicPRL05} considered a stochastic equation without the nonlinear
term $b\rho^2$ and a purely deterministic equation involving the nonlinear term
only, given as
\begin{equation}
\dot\rho = c+\tilde{a}\rho + \sigma \sqrt{\rho}~ \xi~, ~~\dot\rho = - b
\rho^2~. \label{DornicSplitting}
\end{equation}
The FP equation associated with the first stochastic equation can be solved
exactly~\cite{FellerAM51,DornicPRL05}. The conditional probability density
${\cal P}_{\rm S}(\rho, t; \rho_0)$ can be obtained analytically for an initial
value $\rho_0$, vanishing for $\rho<0$ at any time. The deterministic equation
is trivially integrated as $\rho_{\rm D} (t; \rho_0)$, which also preserves the
non-negativity of $\rho$.

The numerical integration during $\Delta t$ is done as follows:
Given an initial value $\rho_t$ at time $t$, $\rho$ is updated
first by sampling a value appropriate to the exact probability
distribution ${\cal P}_{\rm S}(\rho, \Delta t; \rho_t)$, which is
denoted by $\rho_{\rm S}(\Delta t;\rho_t)$. Then, resetting this
updated value as an initial value at time $t$, we integrate the
deterministic equation over $\Delta t$, which yields
\begin{equation}
\rho_{t+\Delta t}=\rho_{\rm D}(\Delta t;\rho_{\rm S}(\Delta
t;\rho_t))~. \label{DornicTE}
\end{equation}
The same procedure is performed for all sites in parallel. This
constitutes a single step in discrete time dynamics with the time
interval $\Delta t$. The next step follows with new initial values
of $\rho_{t+\Delta t}$ and newly tuned values of $c$. This is the
idea introduced in Ref.~\cite{DornicPRL05}.

The splitting scheme can be described mathematically in terms of
the probability density ${\cal P}(\rho,t)$. The probability
density after a single step can be written by ${\cal
P}(\rho,t+\Delta t) = {\cal L}(a,b,D,\sigma) {\cal P}(\rho,t)$,
where ${\cal L}$ is the time evolution operator. Notice that
${\cal L}$ is the product of two consecutive evolution operators,
having the form of $e^{\Delta t L_{\rm D}}e^{\Delta t L_{\rm S}}$,
where $L_{\rm S,D}$ are the Fokker-Planck (FP)
operators~\cite{GardinerHS} associated with the stochastic and
deterministic differential equations, respectively in
Eq.~(\ref{DornicSplitting}). In the Ito calculus, one writes
\begin{equation}
{\cal L} = e^{\Delta t L_{\rm D}}e^{\Delta t L_{\rm S}}=e^{\Delta t\hat
Pb\rho^2}e^{\Delta t(-\hat P(c+\tilde{a}\rho) + \hat P^2 \sigma^2 \rho)}\ ,
\label{DornicFP}
\end{equation}
where $\hat P \equiv \partial/\partial \rho$. We remark that
Eq.~(\ref{DornicFP}) compactly contains the whole information of the
operator-splitting method represented by Eqs.~(\ref{DornicSplitting}) and
(\ref{DornicTE}).

The exact FP operator is given by $L=L_{\rm S} + L_{\rm D}$. Due
to the non-cummutativity of two operators, $[L_{\rm S}, L_{\rm
D}]\neq 0$, the exact evolution operator ${\cal L}_{\rm
exact}=e^{\Delta t L}$ differs from the operator-splitting
evolution operator ${\cal L}$ in higher orders of $\Delta t$. The
difference between ${\cal L}_{\rm exact}$ and $\cal L$ can be
found systematically in power series of $\Delta t$ using the BCH
formula: for $e^Z=e^A e^B$, $Z$ can be expressed as~\cite{HallLLR}
\begin{equation}
Z = A+\int_0^1 dt g\left( e^{{\rm ad}_A}\phantom{i} e^{t\phantom{i}{\rm ad}_B}
\right)B\ , \label{BCH-I}
\end{equation}
where $g(x) \equiv 1+\sum_{m=1}^\infty
{{(-1)^{m+1}}\over{m(m+1)}}(x-1)^m$ and ${\rm ad}_X$ is a linear
map of which the operation is defined by ${\rm ad}_X Y = [X,Y]$.
This leads to the rather familiar BCH formula;
\begin{eqnarray}
Z &=& A+B+{1\over2} [A,B] + {1\over{12}}[A,[A,B]] -
{1\over{12}}[B,[A,B]] \nonumber \\ &&  + .. + k_{w_1,..,w_n}
[w_1,[..[w_n,[A,B]]..]] + ...\ , \label{BCHxy}
\end{eqnarray}
where $w_i$ stands for either $A$ or $B$, and $k_{w_1,..,w_n}$ is a constant
scalar. Here we do not write down the explicit expression of $k_{w_1,..,w_n}$,
but only note that its sign constantly changes with $n$.

With $A=\Delta t L_{\rm D}$ and $B=\Delta t L_{\rm S}$ as in
Eq.~(\ref{DornicFP}), the commutator $[A,B]$ produces a new type
of higher-order noise terms such as $\hat P^2 \rho^2$ by the
interplay of the nonlinear term $\hat P \rho^2$ and the noise term
$\hat P^2 \rho$. Through the nested commutators in
Eq.~(\ref{BCHxy}), the operator-splitting FP operator $\ln {\cal
L}=Z$ includes not only higher order deterministic terms like
$\hat P \rho^n$ but also higher order noise terms like $\hat P^m
\rho^n$. The coefficients of the preexisting lower order terms
such as $\hat P \rho$, $\hat P \rho^2$, and $\hat P^2 \rho$ are
also modified. In the renormalization group sense, the higher
order terms are usually irrelevant to the DP critical behavior,
but only when the appropriate stability condition is satisfied.
For example, the fixed point solution of the deterministic part in
the operator-splitting FP operator should not have a different
structure from that in the exact FP operator. Therefore the
stability condition depends critically on the detailed $\Delta
t$-dependence of the coefficients.  Unfortunately, the complexity
of $k_{w_1,..,w_n}$ do not allow us to judge the stability
criteria in a sensible way. Any conclusion derived from a
truncated finite series in the perturbative expansion of
Eq.~(\ref{BCHxy}) may not contain relevant information on the
stability, especially due to the alternating nature of the series.
It seems also impossible to sum up the infinite series in a closed
form even for the coefficients of the lower order terms. Thus it
is not clear that the modified coefficients due to the
operator-splitting method employed in Ref.~\cite{DornicPRL05}
still guarantee the stability of the DP-type solutions for any
value of $\Delta t$.

Now we focus on classifying various operator-splitting schemes into equivalent
classes, which will greatly reduce the efforts to derive the stability criteria
for $\Delta t$ in general operator-splitting methods. First, we notice that the
nested commutators in Eqs.~(\ref{BCH-I}) and (\ref{BCHxy}) can be easily summed
up when the commutator $[A,B]$ can be written as a linear combination of $A$
and $B$ only. Consider the simple case of $[A,B]=\gamma B$ with a real constant
$\gamma$. As ${\rm ad}_B B =0$, it is easy to show that $g(e^{{\rm ad}_A}
e^{t{\rm ad}_B})B = (1+\sum_{m=1}^\infty {{(-1)^{m+1}}\over{m(m+1)}}(e^\gamma
-1)^m)B$. Hence we obtain
\begin{equation}
Z = A + \alpha_\gamma B, \label{TRC}
\end{equation}
where  $\alpha_\gamma=\gamma/(1-e^{-\gamma})$. Therefore we find
$e^A e^B = e^{A+\alpha_\gamma B}$ or equivalently $e^B e^A =
e^{\alpha_{-\gamma}B+A}$. A general case of $[A,B]=\gamma_A A
+\gamma_B B$ can be reduced to the simple case by replacing $B$ by
${\tilde B}$ proportional to the commutator. The rather
complicated result is not shown here.

Next, we observe that the components of the FP operator $L$
satisfy the above special commutation relation such as $[\hat P
\rho, \hat P] = -\hat P$, $[\hat P \rho, \hat P \rho^2]=\hat P
\rho^2$, and $[\hat P \rho, \hat P^2 \rho]=-\hat P^2 \rho$. This
implies that the linear term $\hat P\rho$ can move around rather
freely between $L_{\rm S}$ and $L_{\rm D}$ without causing too
much complication in the operator-splitting FP operator $Z$. We
note that the other commutators do not satisfy the special
relation and the algebra is not closed.

Consider a different splitting scheme where the linear term is
included in the deterministic part. The splitted Langevin
equations are
\begin{equation}
\dot\rho = c+\sigma \sqrt{\rho}~ \xi~, ~~\dot\rho = \tilde{a}\rho - b \rho^2~.
\label{testSplitting}
\end{equation}
As in the previous splitting in Eq.~(\ref{DornicSplitting}), both
equations can be treated exactly. The corresponding evolution
operator can be written as
\begin{equation}
{\cal L}'(a,b,D,\sigma) = e^{\Delta t\hat P(b\rho^2-\tilde
a\rho)}e^{\Delta t(-\hat P c + \hat P^2 \sigma^2 \rho)}.
\label{testFP}
\end{equation}
Using the identity in Eq.~(\ref{TRC}), we split the first exponential map as
\begin{equation}
{\cal L}'(a,b,D,\sigma) = e^{\Delta t\hat P (b/\alpha_{\tilde a \Delta
t})\rho^2} e^{-\Delta t \hat P \tilde a\rho} e^{\Delta t(-\hat P c + \hat P^2
\sigma^2 \rho)}\ . \label{TR-1}
\end{equation}
The last two exponential maps can be merged together as
\begin{equation}
{\cal L}'(a,b,D,\sigma) = e^{\Delta t\hat P (b/\alpha_{\tilde a
\Delta t})\rho^2} e^{\Delta t(-\hat P \tilde a \rho -
\alpha_{\tilde a \Delta t}(\hat P c - \hat P^2 \sigma^2 \rho))}.
\label{TR-2}
\end{equation}

By comparing Eqs.~(\ref{DornicFP}) and (\ref{TR-2}), one establishes the
relation between the two operator-splitting schemes as
\begin{equation}
{\cal L}'(a,b,D,\sigma) = {\cal L} (a',b',D',\sigma')~,
\label{CTR-FP}
\end{equation}
where
\begin{eqnarray}
&& a'=a+2dD(\alpha_{\tilde a \Delta t} -1)/(\Delta x)^2\
,~b'=b/\alpha_{\tilde a \Delta t}\ , \nonumber
\\
&& D'=\alpha_{\tilde a \Delta t} D\ , ~\sigma'= \sqrt{\alpha_{\tilde
a \Delta t}} \sigma \ . \label{CTR}
\end{eqnarray}
Since $\alpha$ is positive-definite and does not vanish nor
diverge for any finite $\tilde a \Delta t$, the transformation
between ${\cal L}$ and ${\cal L}'$ forms an {\it automorphism}
(one-to-one and onto itself) in the parameter space of
$(a,b,D,\sigma)$. The transformation preserves the sign of the
parameters $(b,D,\sigma)$ except $a$ (tuning parameter) where the
reformulation of the discrete Laplacian is involved. Therefore,
the different operator-splitting methods are related to each other
only by trivial rescaling of parameters with a shift of the
critical point. Any phenomenon observed in one splitting scheme is
also expected in the other scheme and the stability conditions for
$\Delta t$ can be traced using the transformation of
Eq.~(\ref{CTR}) if it is known for one specific operator-splitting
method.

We may consider a more general splitting where the linear term is arbitrarily
divided into two parts. That is,  the splitted Langevin equations are now
\begin{equation}
\dot\rho = c+(\tilde a - k)\rho +\sigma \sqrt{\rho} ~\xi~, ~~\dot\rho = k \rho
- b \rho^2~. \label{testSplitting2}
\end{equation}
where $k$ is a real arbitrary constant. Note that we can still integrate both
equations exactly. The corresponding time evolution operator is given as
\begin{equation}
{\cal L}_k (a,b,D,\sigma) = e^{\Delta t\hat P(b\rho^2 - k\rho)}e^{\Delta
t(-\hat P (c + (\tilde a - k)\rho) + \hat P^2 \sigma^2 \rho)}. \label{GFP}
\end{equation}
Similarly, we  find
\begin{equation}
{\cal L}_k(a,b,D,\sigma) = e^{\Delta t\hat P (b/\alpha_{k \Delta
t})\rho^2} e^{\Delta t (-\hat P \tilde a \rho + \beta_k (-\hat P c +
\hat P^2 \sigma^2 \rho))}, \label{GFP-f}
\end{equation}
where
\begin{equation}
\beta_k = \alpha_{\tilde a\Delta t} \alpha_{(\tilde a - k)\Delta
t}^{-1}={{\tilde a}\over{\tilde a - k}} {{e^{\tilde a \Delta t}-e^{k
\Delta t}}\over{e^{\tilde a \Delta t} - 1}} \ . \label{beta}
\end{equation}

Hence one can generalize Eqs.~(\ref{CTR-FP}) and (\ref{CTR}) as
follows;
\begin{equation}
{\cal L}_k(a,b,D,\sigma) = {\cal L} (a_k,b_k,D_k,\sigma_k)~,
\label{GCTR-FP}
\end{equation}
where
\begin{eqnarray}
&& a_k=a+2dD(\beta_k -1)/(\Delta x)^2\ ,~b_k=b/\alpha_{k \Delta t}\
, \nonumber
\\
&& D_k=\beta_k D\ , ~\sigma_k = \sqrt{\beta_k} \sigma \ .
\label{GCTR}
\end{eqnarray}
Due to the property of $\beta$, inherited from that of $\alpha$, the
transformation between ${\cal L}_k$ and ${\cal L}$ also forms an automorphism
for any $k$. Consequently, the solution-structure yielded by ${\cal L}_k$ is
always preserved irrespective of $k$, and thus such operations can be
represented by ${\cal L}$. This directly demonstrates that ${\cal L}_k$'s form
an equivalent class of operator-splitting integration of the DP Langevin
equation.

In summary, we present a compact mathematical description of the so-called
operator-splitting method, which was proposed in
Refs.~\cite{PechenikPRE99,DornicPRL05} for the numerical integration of the DP
Langevin equation of Eq.~({\ref{dple}). Based on this, we show analytically
that some splitting methods are mathematically equivalent with the explicit
transformation function of the model parameters.  Consequently, we find that
the splitting methods ${\cal L}_k$'s form an equivalent class of integration in
the sense that the solution-structure and the transition property between the
solutions are always conserved for any $k$. In the meantime, we also address
that the difference between the original dynamics and that by the
operator-splitting scheme is traceable by the perturbation theory of the
presented mathematical description. However, the information from the
perturbation theory seems not sufficient to decide whether the splitting scheme
still preserves the essential feature of the original dynamics. Nonetheless,
our work on the equivalence class will be of considerable help to examine the
validity of the operator-splitting scheme in studying the universality of the
DP Langivin equation.




\end{document}